\documentclass{mn2e}
\usepackage{epsfig}
\usepackage{epsf}
\usepackage{graphicx}

\title[Equatorial mass loss and X-rays from NaSt1]{Multiwavelength observations of NaSt1 (WR\,122): Equatorial mass loss and X-rays from an interacting Wolf-Rayet binary}

\author[Jon Mauerhan et al.]{Jon Mauerhan$^{1}$\thanks{E-mail:
mauerhan@astro.berkeley.edu}, Nathan Smith$^{2}$, Schuyler D. Van Dyk$^{3}$, Katie M. Morzinski$^{2}$, \newauthor Laird M. Close$^{2}$, Philip M. Hinz$^2$, Jared R. Males$^{2}$, and Timothy J. Rodigas$^{4}$ \\
$^{1}$Department of Astronomy, University of California, Berkeley, CA 94720-3411, USA \\
$^{2}$Steward Observatory, University of Arizona, 933 N. Cherry Ave., Tucson, AZ 85721, USA \\
$^{3}$IPAC/Caltech, Mail Code 100-22, Pasadena, CA 91125, USA \\
$^{4}$Department of Terrestrial Magnetism, Carnegie Institute of Washington, 5241 Broad Branch Road, NW, 
Washington, DC 20015, USA}
 
\begin{document}
\date{Accepted 0000, Received 0000, in original form 0000}
\pagerange{\pageref{firstpage}--\pageref{lastpage}} \pubyear{2002}
\def\arcdeg{\degr}
\maketitle
\label{firstpage}

\begin{abstract}
NaSt1 (aka Wolf-Rayet 122) is a peculiar emission-line star embedded in an extended nebula of [N~{\sc ii}] emission with a compact dusty core. This object was characterized by Crowther \& Smith (1999) as a Wolf-Rayet (WR) star cloaked in an opaque nebula of CNO-processed material, perhaps analogous to $\eta$ Car and its Homunculus nebula, albeit with a hotter central source.  To discern the morphology of the [N~{\sc ii}] nebula we performed narrowband imaging using the \textit{Hubble Space Telescope} and Wide-field Camera 3. The images reveal that the nebula has a disk-like geometry tilted $\approx$12$^{\circ}$ from edge-on, composed of a bright central ellipsoid surrounded by a larger clumpy ring.  Ground-based spectroscopy reveals radial velocity structure ($\pm10$\,km\,s$^{-1}$) near the outer portions of the nebula's major axis, which is likely to be the imprint of outflowing gas.  Near-infrared adaptive-optics imaging with Magellan AO has resolved a compact ellipsoid of $K_s$-band emission aligned with the larger [N\,{\sc ii}] nebula, which we suspect is the result of scattered He\,{\sc i} line emission ($\lambda$2.06\,{$\mu$m}). Observations with the \textit{Chandra X-ray Observatory} have revealed an X-ray point source at the core of the nebula that is heavily absorbed at energies $<1$\,keV and has properties consistent with WR stars and colliding-wind binaries. We suggest that NaSt1 is a WR binary embedded in an equatorial outflow that formed as the result of non-conservative mass transfer. NaSt1 thus appears to be a rare and important example of a stripped-envelope WR forming through binary interaction, caught in the brief Roche-lobe overflow phase.
 \end{abstract}


\section{INTRODUCTION}
The physical processes by which massive stars shed their hydrogen envelopes and become Wolf-Rayet (WR) stars present an interesting challenge to the standard stellar evolutionary paradigm (see review by Smith 2014 and references therein). Mass loss via steady stellar winds is punctuated by discrete, episodic, and sometimes explosive mass ejections that occur during the post main-sequence, in the so-called Luminous Blue Variable (LBV) phase (Smith \& Owocki 2006). The physics that trigger and drive these intense bouts of eruptive mass loss are still not understood, yet this mode is suspected to be a key channel by which H-depleted WRs form. In addition to discrete eruptions, binary mass transfer via Roche lobe overflow (RLOF) is another very important (perhaps even dominant) channel for producing stripped-envelope WRs, and may even be critical to account for Type Ib/c supernova progenitors (Smith et al. 2011a; Dessart et al. 2012; Hachinger et al. 2012). Since the time-scale of eruptive mass loss in such systems is brief, only a precious few Galactic examples exist, especially for cases where stellar multiplicity (i.e., binary vs. single) is an influential factor. A particularly interesting example is RY Scuti -- a massive binary undergoing mass transfer that recently resulted in an intense bout of equatorial mass loss, and where one of the stellar components is likely on its way to becoming a stripped-envelope WR (Smith et al. 2011b).

{NaSt1}, aka WR\,122, is an evolved massive star that has defied characterization. It was discovered by Nassau \& Stephenson (1963), who proposed a WR classification based on the strong emission-line spectrum.  It was later reclassified as a late nitrogen-type WR (WN type) based on nitrogen emission-line ratios in the optical spectrum (Massey \& Conti 1983), then later proposed to be a B[e], O[e], or Ofpe/WN9 star based on its optical--IR spectral properties (van der Hucht 1997). Subsequently, more detailed multi-wavelength investigation by Crowther \& Smith (1999; hereafter CS99) showed that although NaSt1 is indeed a hot luminous object with $L\sim10^{5.1}$--$10^{6.1}~L_{\odot}$, its emission-line spectrum is mostly of nebular origin, not stellar. Moreover, ground-based narrowband imaging through an [N~{\sc ii}] filter centered near wavelength $\lambda$6584\,{\AA} revealed the presence of a compact circumstellar nebula with a diameter of $\approx$7{\arcsec}, whose elongated yet poorly resolved dimensions gave the impression of possible bipolar morphology. No comparable extended component was inferred for H~{\sc i} gas based on the much weaker detection of nebulosity in images through a broader H$\alpha+$[N~{\sc ii}] filter (which could simply be attributable to [N~{\sc ii}] emission in the passband). He {\sc i} and He {\sc ii} narrowband images revealed no extended component either. 

Derivation of the chemical abundances in NaSt1's nebula via quantitative spectroscopy by CS99 showed it to be extremely nitrogen enriched yet carbon and oxygen deficient, which indicates that the nebula consists of highly CNO-processed material ejected from an evolved massive star. CS99 pointed out that the only other massive star {\it known} to be enshrouded in a nebula having a comparable level of nitrogen enrichment is the LBV$+$OB binary $\eta$ Car and its Homunculus nebula (Davidson et al. 1986; Dufour et al. 1997; Smith \& Morse 2004). However, CS99 note that the chemical abundance ratios in the NaSt1 nebula are characteristic of a WR, and probably inconsistent with a red supergiant or H-rich LBV. Moreover, the derived ionization potential requires that the central ionizing source be $T>>30$\,kK, which suggests a spectral subtype earlier than WN6--7. The absence of the characteristic broad emission lines from a WR wind, however, implies that the stellar system is deeply embedded in an opaque nebula. High-resolution spectroscopy by CS99 revealed that the nebular lines have a double-peaked morphology, which suggests complex geometry for the outflow.

\begin{figure*}
\includegraphics[width=6.9in]{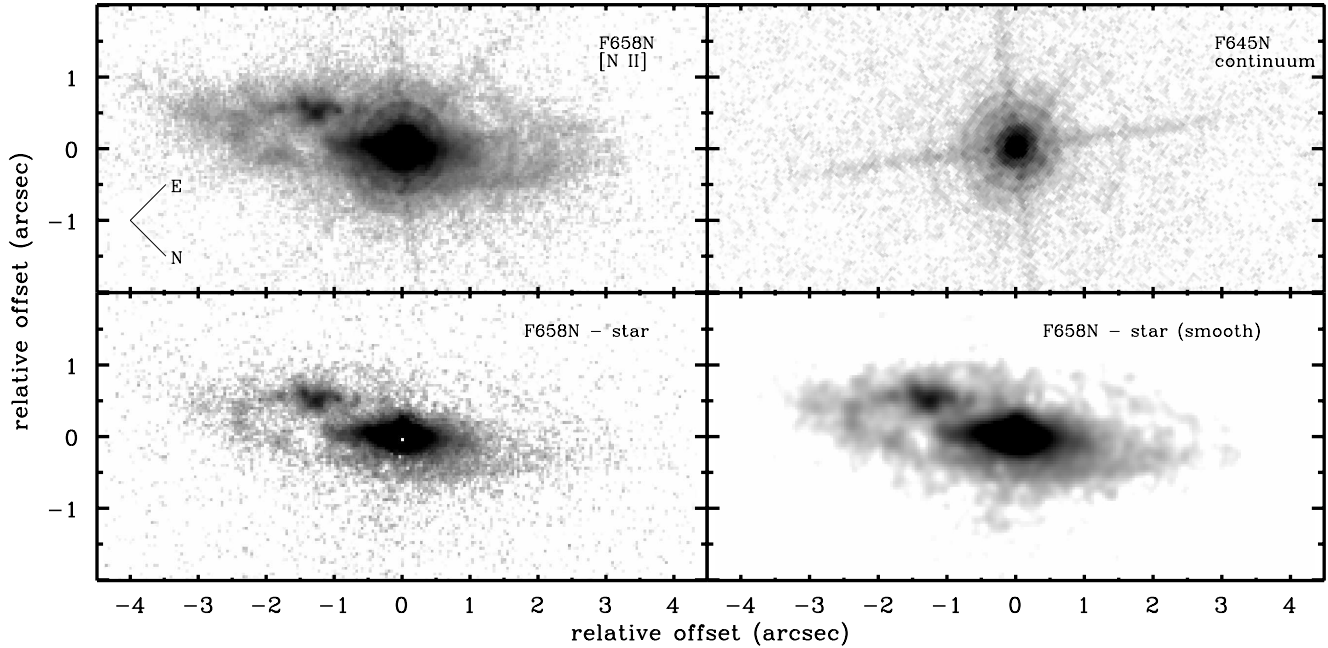}
\caption{{\textit{HST}}/WFC3 narrowband images of NaSt1 through F658N and F645N filters (logarithmic stretch), corresponding to the [N~{\sc ii}] emission line and adjacent continuum, respectively (upper panels). The lower panels show the F658N image with the central point source removed, and a 3-pixel smoothed version.}
\label{fig:hst_ims}
\end{figure*}

NaSt1 also exhibits strong thermal excess from circumstellar dust, but from a much more compact region than the [N~{\sc ii}] nebula. CS99 presented evidence for a two-temperature spectral energy distribution in the infrared, consisting of warm ($T\approx700$~K) and hot ($T\approx2000$~K) components. A mid-infrared spectrum by Smith \& Houck (2001) showed it to be mostly featureless in the 8--14\,$\mu$m wavelength range, with the exception of two narrow nebular lines of [Fe\,{\sc ii}] and [Ne\,{\sc ii}]. The results of mid-infrared interferometric imaging at $\sim10$\,{$\mu$m wavelength by Rajagopal et al. (2007) were consistent with at least two infrared components having angular sizes within 50--60 mas. Based on the likely distance range of 1--3 kpc and on the measured outflow velocities from CS99, the results of Rajagopal et al. (2007) imply that the dust originates from within $<$100 AU of the central source\footnote{Note, however, that sensitivity of interferometry to spatial frequencies depends on the distribution of the baselines.} and thus has a short dynamical timescale of only 30 yr, which implies only very recent and likely ongoing dust formation.  By comparison, the outer [N~{\sc ii}] nebula has a much larger size of $\approx$7{\arcsec} (6,800--22,400 AU). The presence of an additional component of faint thermal dust emission associated with the extended nebular structure, to our knowledge, has never been assessed with sufficient spatial resolution.    

The collective properties of NaSt1 led CS99 to conclude that this object could be a WR cloaked in an $\eta$ Car-like ejection nebula, while the interferometry of Rajagopal et al. (2007) motivates a search for extended structure in the near-infrared. Here, we investigate the situation further by obtaining high-resolution narrowband images of NaSt1's nebula in the light of [N~{\sc ii}] using the \textit{Hubble Space Telescope} ({\textit{HST}), and adaptive-optics (AO) imaging in the near-infrared with Magellan AO. Moreover, the possible presence of a WR star in an $\eta$ Car-like system also warrants the search for an X-ray counterpart to NaSt1, which could be expected to arise from shocks produced by hypersonic wind flows that are often associated with WRs. NaSt1 was not detected by the ROSAT all-sky Survey, but this is not particularly surprising, given the relatively shallow sensitivity of this survey, in conjunction with the estimated distance to the source and the anticipated absorption to X-rays with energies $\le1$--2\,keV from interstellar material. Thus, we have obtained deeper X-ray imaging observations of NaSt1 using the \textit{Chandra X-ray Observatory}. The resulting data from these space-based and AO facilities provide new important clues for the nature and evolutionary status of this object. 

\section{OBSERVATIONS AND RESULTS}
\subsection{Narrowband [N~{\sc ii}] imaging with {\textit{HST}}/WFC3}
NaSt1 was first observed with {\textit{HST}} on 2013 March 20 (UT 02:34) using the Wide-field and Planetary Camera 3 (WFC3). Five 176\,s exposures were obtained through both the F658N and F645N filters, which have peak transmission wavelengths centered on the [N~{\sc ii}] $\lambda$6584~{\AA} line and the adjacent continuum, respectively. A five-point sub-pixel dither pattern (WFC3-UVIS-DITHER-BOX) was executed between individual exposures to improve sampling of the WFC3 point spread function (PSF). Since NaSt1 exhibits a small angular size, images were read out in a 1K$\times$1K subarray mode to decrease readout overhead. To mitigate the effect of degrading charge transfer efficiency (CTE) in the WFC3 CCD chip, a 12 $e^-$ post-flash was performed at the end of each exposure, which slightly elevated the background counts and mitigated against CTE losses.
 
Some of the F658N and F645N images obtained on March 20 were severely degraded because the fine-guidance sensors failed to track the guide stars from the Guide Star Catalog, which turned out to be blends of multiple stars at {\textit{HST}} resolution. The observations thus defaulted to GYRO pointing control and guiding, with no fine guidance.  The resulting stellar PSFs, which normally have optimal full-width at half-maximum (FWHM) values of $\sim0\farcs07$, were highly elliptical and, in some cases, double-peaked, exhibiting FWHM values ranging from $\sim{0\farcs10}$ to $\sim{0\farcs19}$. Although several of the individual images from this visit were salvageable and scientifically useful, the overall telescope drift during the exposures significantly limited our planned imaging sensitivity and resolution. NaSt1 was thus re-observed on 27 April 2013 (UT 21:04), using an observing strategy identical to the 20 March observations, although with new guide stars that were confirmed to be single. The repeated observations achieved the intended sub-pixel dithering accuracy and yielded images with FWHM PSF cores of $0\farcs09$ using the \textit{Multidrizzle} pipeline. Three of the lesser quality F658N [N~{\sc ii}] images and two of the F654N images from the initial observations were of acceptable quality and worth adding into our final drizzled image. 

\begin{figure*}
\includegraphics[width=6.9in]{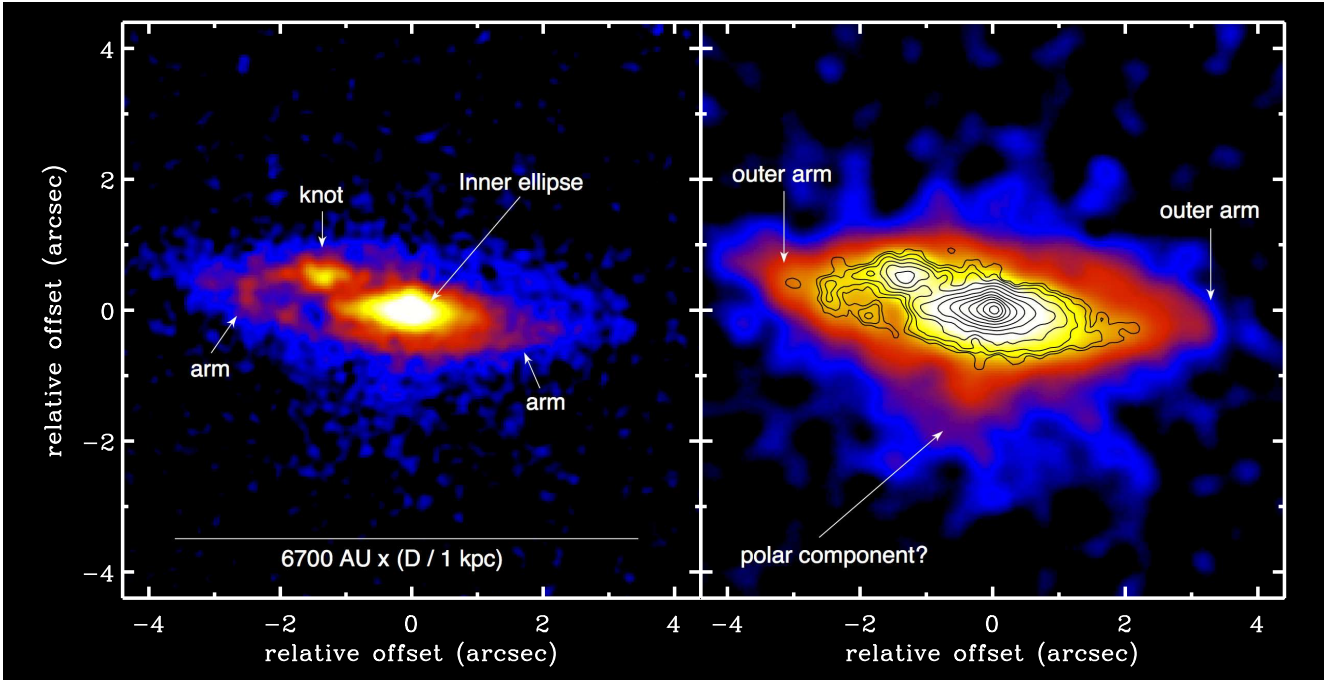}
\caption{\textit{Left}: star-subtracted 3-pixel smoothed F658N image of NaSt1 (logarithmic stretch), displayed in false color to enhance the nebular structure. \textit{Right}: same as above with 9-pixel smooth and more extreme stretch to enhance the faint outer nebula. Flux contours of the brighter central regions from the 3-pixel smoothed image are overlaid for reference. The orientation of the images is the same as in Figure\,\ref{fig:hst_ims}.}
\label{fig:im_smooth}
\end{figure*}

Figure\,\ref{fig:hst_ims} shows the \textit{drizzled} F658N and F654N images of NaSt1. Two sets of diffraction spikes with different intensities can be seen, as a result of combining images from two separate visits with {\textit{HST}} at different roll angles for the spacecraft. The nebula is clearly discernible in the F658N image, while no extended structure is immediately apparent in the F654N image. Subtraction of the bright central point source was challenging, and the most optimal strategy for getting the best possible subtraction was difficult to converge upon. The F658N line and F654N continuum filters produce PSFs that have substantially different speckle distributions and Airy-ring patterns, so simple subtraction of the continuum image from the line image did not produce satisfactory results, leaving large residuals with complex mottled structure. Furthermore, both filters exhibit unique ring-like ghost patterns that extend to $\approx0\farcs9$ from the PSF centroid\footnote{WFC3 Instrument Handbook (\S6.5.3).}. Subtraction of artificial PSFs generated by the \textit{Tiny Tim} software also produced unsatisfactory results, since the PSF models do not account for filter-dependent ghost artifacts. We thus found that the best strategy for subtracting the central point source was to use a bright neighboring star from the same images, 15{\arcsec} away from NaSt1. We cut out a sub-image centered on this star, smoothed the pixels outside of the PSF halo to mitigate the effect of added noise to the subtracted image, and then scaled the `model'  incrementally to find the optimal subtraction that minimized the difference residuals. In the end, the best scale factor of 0.8 removed all the higher-order residual features in the subtracted image very well (the Airy rings, speckles, and diffraction spikes), but left a substantial amount of point-like [N\,{\sc ii}] emission at the core of the nebula. Some of this point-like emission could be real, or could also be an artifact of imperfect subtraction. To check for the presence of compact continuum emission, we performed the same strategy for the F645N image and determined that there is no detection of an extended component from continuum emission or scattered light at the image resolution, just noise residuals. \medskip

{\noindent \bf Equatorial geometry}\smallskip \\ The lower panels of Figure\,\ref{fig:hst_ims} show the star-subtracted F658N image and a 3-pixel smoothed version, which shows the inner structure of the nebula more clearly. Overall, NaSt1's nebula exhibits a complex multi-component structure having disk-like morphology with approximate major and minor axes of 7{\arcsec} and 2{\arcsec}, respectively, corresponding to an axis ratio $\approx$3.5. The emitting material thus appears to have equatorial distribution and, if approximately circular, is tilted at an angle of $\approx12\pm3^{\circ}$ from edge-on (inclination angle of $i=78\pm3^{\circ}$). The central point source is slightly offset by 60\,mas from the geometric center of the nebula, as determined by the distribution of isophotal contours of the faintest outer regions.  Figure\,\ref{fig:im_smooth} shows smoothed false-color versions of the star-subtracted [N~{\sc ii}] image. The inner portion of the nebula is composed of a relatively bright and smooth central ellipsoid with major and minor axes of 2\farcs3 and 0\farcs6, respectively, consistent with the same axis ratio as the greater structure. Based on the distribution of isophotal contours, the central point source also appears to be slightly offset from the center of this inner ellipse by 50\,mas. Outside of the central ellipse there is a gap $\approx$1{\arcsec} in length along the major axis, most notably on the south-west side. Beyond this gap, the outer nebula is composed of a clumpy ring or arm that can possibly be traced around the entirety of the structure. There is a particularly bright cluster or knot of emission clumps within this outer structure, at $\approx$1\farcs4 south/south-west of the central point source. The brightest regions of the outer ring on either side of the center appear to have approximate point-reflection symmetry, which hints at a spiral structure for the outer ring/arm(s). The offset of the central point source from the apparent geometric center of the nebula could also be an indication of a one-armed spiral morphology. A highly contrast-stretched 9-pixel smoothed image (Figure\,\ref{fig:im_smooth}, right panel) shows faint nebular structure with a full diameter of 7\farcs7.

 \begin{figure*}
\includegraphics[width=6.9in]{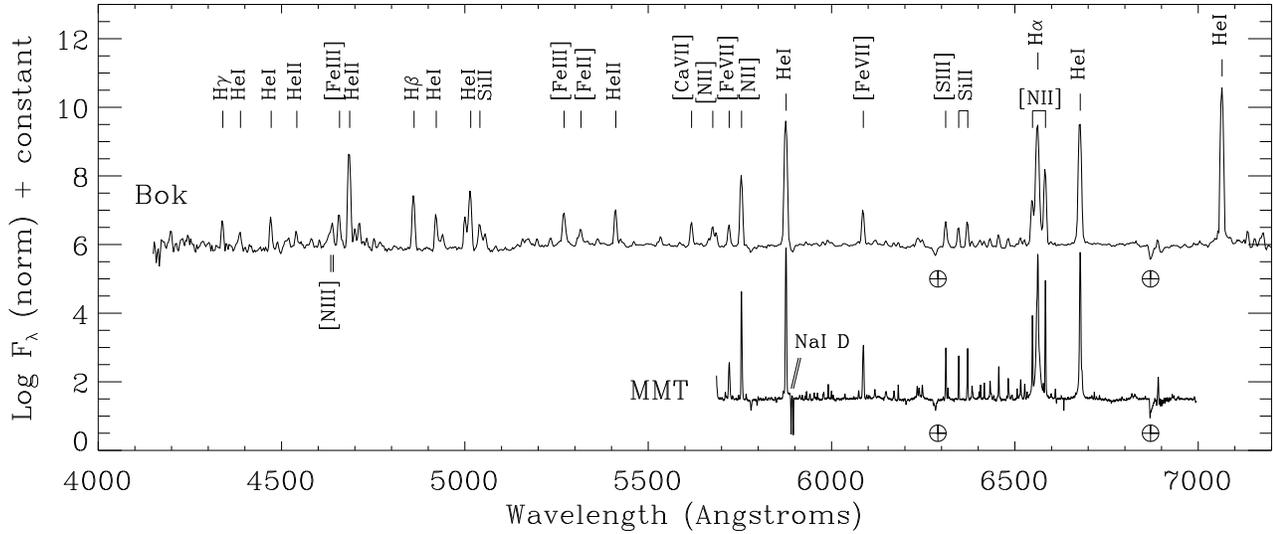}
\caption{Optical spectra of NaSt1 from Bok (upper spectrum) and MMT (lower spectrum). The most prominent nebular features have been marked (see CS99 for a comprehensive list of all emission lines).}
\label{fig:spectra}
\end{figure*}

 \begin{figure*}
\includegraphics[width=6.9in]{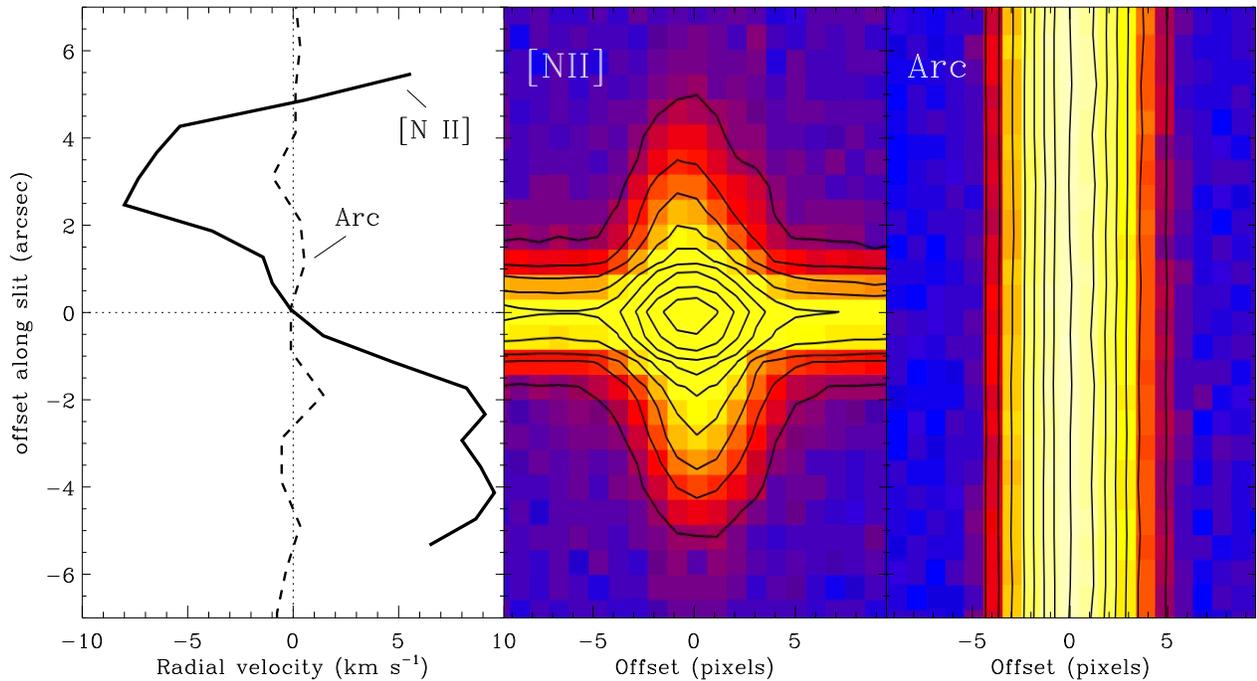}
\caption{Radial velocity of the [N~{\sc ii}] $\lambda$6584 profile (left panel, solid curve) as a function of position along the slit (middle panel), which was approximately aligned with the major axis of the nebula, and the same for an arc line of Ne~{\sc i}\,$\approx$\,6599~{\AA} (right panel). Real kinematic motion for the outer nebula of NaSt1 is apparent. Note that velocity measurements for offsets $<2${\arcsec} are dominated by emission from the central point source, and are not reliable.}
\label{fig:spatial_spec}
\end{figure*}

\begin{figure}
\includegraphics[width=3.3in]{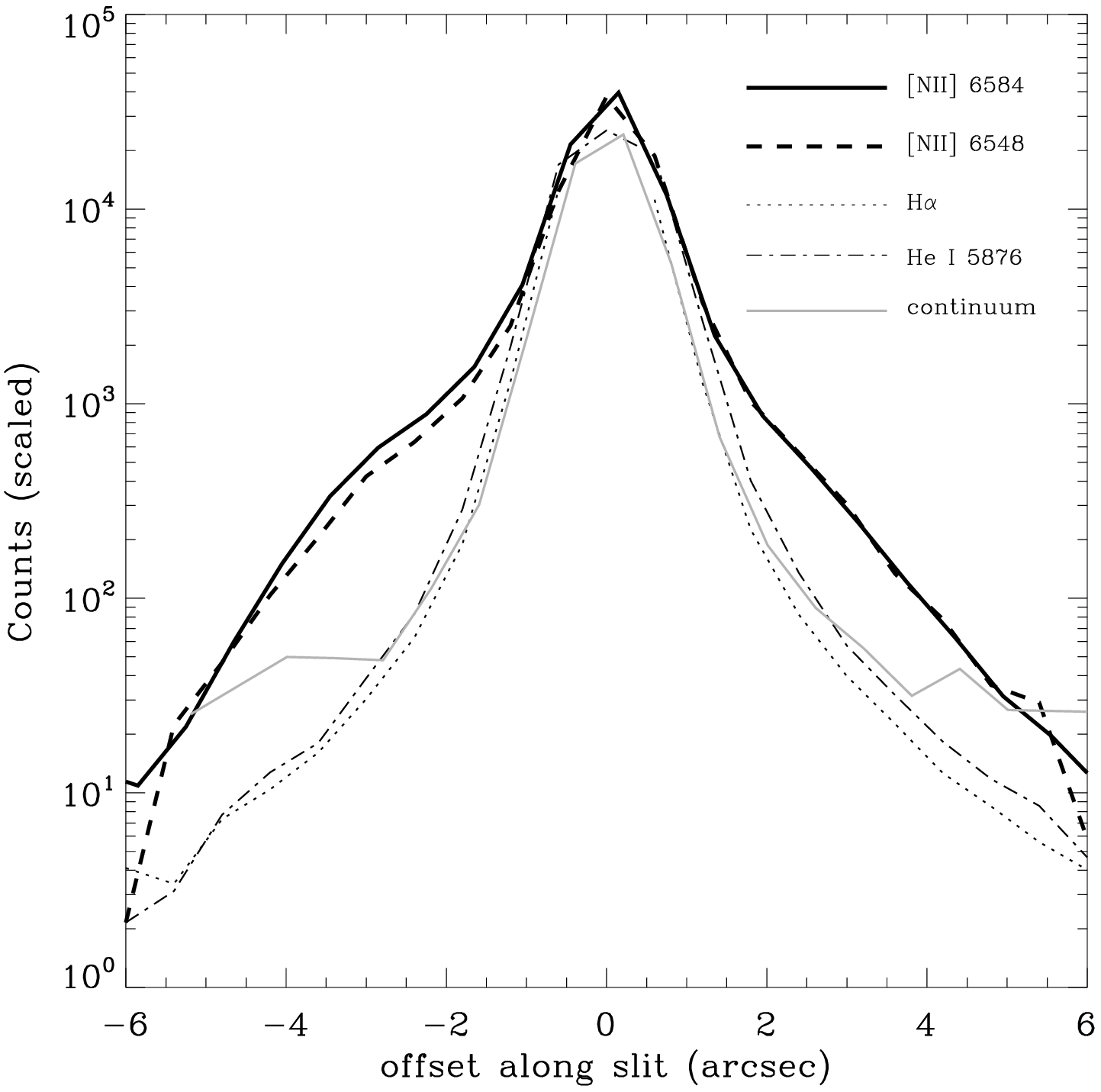}
\caption{\textit{Top}: spatial profile of selected emission lines along the spectrograph slit, which was tilted near the long axis of the nebula at 35$^{\circ}$. The lines are scaled in intensity to roughly match the peak of the [N~{\sc ii}] $\lambda$6584 feature. The nebula is clearly resolved in the [N~{\sc ii}] transitions. The bumps on either side of the continuum profile are an artifact of the scaling up of the continuum, and are not real resolved continuum structure.}
\label{fig:profile}
\end{figure}

\subsection{Optical spectroscopy}
Optical spectroscopy of NaSt1 was performed on 2012 March 31 using the 2.3\,m Bok Telescope at Kitt Peak, Arizona. A low-resolution grating having 400\,l\,mm$^{-1}$ was used in conjunction with a 1\farcs5 slit to achieve a FWHM spectral resolution of 4.6\,{\AA} and a wavelength range of $\lambda\lambda4150$--7200\,{\AA}. Four exposures of 120\,s each were obtained at an airmass of 1.5--1.6 and median combined into a final spectrum. Flat fielding and wavelength calibration were performed by taking spectral images of the internal continuum and arc lamps at the beginning of the night. Flux calibration was performed using spectra of the white dwarf standard star BD$+$42\,2811. Spectroscopy was not performed with the slit aligned to the parallactic angle, so some loss of blue light is expected. All image and spectral reductions were conducted using standard {\sc iraf} routines.

Spectroscopy of NaSt1 was also performed on 2012 Oct 17 with the 6.5\,m telescope at MMT Observatory using the Bluechannel spectrograph. The 1200~l\,mm$^{-1}$ grating was utilized with a 1{\arcsec} slit to achieve a dispersion of $\approx$0.5\,{\AA} per pixel ($\approx$20\,km\,s$^{-1}$ pixel$^{-1}$). For this observation, the slit was oriented at 31$^{\circ}$ from north, approximately parallel to the major axis of the nebula. This strategy was used in order to search for kinematic features that might be discernible along the resolved axis of the nebula. Three exposures of 300 s each were obtained at an airmass of 1.4 and median combined into a final spectrum. Flat fielding and wavelength calibration were performed by taking spectral images of the internal continuum and arc lamps at the beginning and end of the night. Flux calibration was performed using spectra of the white dwarf standard star Feige\,34. Since the slit was not aligned with the parallactic angle, some loss of blue light is expected.

Figure\,\ref{fig:spectra} shows the low- and moderate-resolution optical spectra of NaSt1 from Bok and MMT. The spectral morphology is mostly nebular in nature, exhibiting a complex mix of permitted and forbidden emission lines spanning a broad range of ionization. The comparatively higher resolution spectrum from MMT shows the resolved Na\,{\sc i}\,D doublet from interstellar (and, perhaps, circumstellar) absorption along the line of sight. The MMT spectrum also shows extended Lorentzian wings for the H$\alpha$, close examination of which shows a full possible extent of at least $\approx \pm2000$\,km\,s$^{-1}$ for the wing edges; this Lorentzian profile indicates that H$\alpha$ is not entirely nebular in origin, and suggests that some of the flux is from a physically distinct region of high optical depth, where electron scattering could be occurring. Narrower wings might also be associated with the He\,{\sc} $\lambda$6680\,{\AA} profile, but not at the level observed for H$\alpha$. Overall, the optical spectra presented here are consistent with the previous observations of CS99, and we find no substantial spectral differences between the epochs presented here or with respect to the measurements in the literature. \medskip

{\noindent \bf Spatially resolved kinematics}\smallskip\\
Figure\,\ref{fig:spatial_spec} (center panel) shows a two-dimensional spectral image of the [N\,{\sc ii}] $\lambda$6584 line from the MMT observation, for which the slit was aligned along the major axis of the nebula. A similar image of a neighboring arc line (right panel) from the calibration exposures is also shown for comparison. The left panel of  Figure\,\ref{fig:spatial_spec} shows the measured shift in radial velocity of [N\,{\sc ii}] along the spatial axis, which exhibits an `S' shaped curve having a peak velocity of $\pm$10~km\,s$^{-1}$ at $\approx$3{\arcsec} from the center of the continuum dispersion axis (we note that the velocity measurements for offsets $<2${\arcsec} are dominated by emission from the PSF of the central point-source, and are thus not reliable tracers of the nebular velocity at those positions). We do not see any such velocity structure associated with H$\alpha$ or He\,{\sc i}; this is to be expected, since these features are spatially unresolved. For example, Figure\,\ref{fig:profile} shows the spatial profile of [N\,{\sc ii}] along the slit compared to H$\alpha$, He\,{\sc i}, and the continuum adjacent to [N\,{\sc ii}] $\lambda$6584. It is apparent that the [N\,{\sc ii}] emission from the nebula is resolved, while the other prominent lines and the continuum are not. 

It thus appears likely that the [N~{\sc ii}] emission exhibits real kinematic structure that is intrinsic to the object, likely to be an imprint of outflowing gas. If the nebula is an expanding disk or ring then the relative motions of gas along the slit, which are aligned to the major axis of the nebula, will be mostly transverse, not radial. In this case, the true expansion velocity could be significantly higher. Spectroscopy with high spatial and spectral resolution, sampled along the \textit{minor} axis of the nebula is necessary to gauge the true velocity structure. 

\section{Infrared imaging and Spectroscopy}
\subsection{AO imaging with MagAO$+$Clio2}Near-infrared AO images were obtained on 2013 April 7 with the Magellan 6.5\,m Clay telescope at Las Campanas Observatory during commissioning of the MagAO and Clio2 instrument (Morzinski et al. 2014). The wide camera was used, yielding a plate scale of 27.5\,mas and a field of view of $28\arcsec \times 14\arcsec$. NaSt1 itself served as the natural guide star for the AO system. Thirty images of 40 coadds each were obtained through a $K_s$ filter and 10 images of 40 coadds each were obtained through $L^{\prime}$ and $M^{\prime}$ filters. The $L^{\prime}$ and $M^{\prime}$ observations also included a neutral-density (ND) filter to prevent saturation of the detector.  We note that the $K_s$ filter setup used throughout the observing run did not have a long-wavelength blocker, which resulted in a small amount of leakage of light from 3--5\,{$\mu$m} wavelengths (hereafter, reference to the $K_s$ band includes the potential for some long-wavelength leakage). The telescope was nodded by 4{\arcsec} between exposures for background  subtraction. No flat-fielding was performed. The background-subtracted data were then median combined into final image frames for each filter. Three field stars in the Clio2 image have entries in the United Kingdom Infrared Deep Sky Survey (UKIDSS; Lawrence et al. 2007) Galactic Plane Survey (GPS) catalog (Data Release 7), and were used to calibrate the astrometry. Two of those stars were also present in the {\textit{HST}}/WFC3 frame, which enabled us to align the astrometry of the optical and infrared images to a relative accuracy of  0\farcs064 in physical arcsec.  

Figure\,\ref{fig:clio} (upper three panels) shows the MagAO/Clio2 images of NaSt1 in the three infrared bands. At $M^{\prime}$ NaSt1 appears to be unresolved, as the diffraction-limited structure of the circular PSF is clearly discernible in the form of multiple concentric Airy rings. The FWHM of the $M^{\prime}$ PSF centroid is 5.96 pixels, which corresponds to an angular size of 164\,mas. At $L^{\prime}$ the FWHM is 142\,mas and the first Airy ring is marginally discernible. The PSF halo is asymmetrically bright, which could indicate that the source is marginally resolved. The $K_s$-band image, by comparison, shows that the source is fully resolved into an ellipsoidal structure that is aligned with the halo asymmetry observed at $L^{\prime}$. The FWHM of the $K_s$ centroid along the longest axis is 171\,mas, substantially broader than the PSF centroids of the $L^{\prime}$ and $M^{\prime}$ images. It thus seems unlikely that the ellipsoidal structure in the $K_s$ band has resulted from the small amount of potential leakage from 3--5\,{$\mu$m} light.  Moreover, observations of an unresolved star (2MASS\,J12073100$-$3932281) having the same Strehl ratio produced a substantially more compact stellar $K_s$ PSF having FWHM=123\,mas. In addition, observations of HD\,73634, which is comparably bright to NaSt1 in the 3--5\,{$\mu$m} range, produced a stellar PSF having FWHM=126\,mas. Any potential degradation of the stellar PSF from long-wavelength leakage, if present, should have been noticeable in the images of this star as well. No such effect is observed, and we thus conclude that the extended $K_s$-band emission around NaSt1 is real. At $L^{\prime}$ and $M^{\prime}$, however, NaSt1 has the same FWHM size as the comparison stars and thus appears to be unresolved at these longer wavelengths, although it is possible that there is marginally resolved structure in the halo of the $L^{\prime}$ source.

\begin{figure}
\includegraphics[width=3.3in]{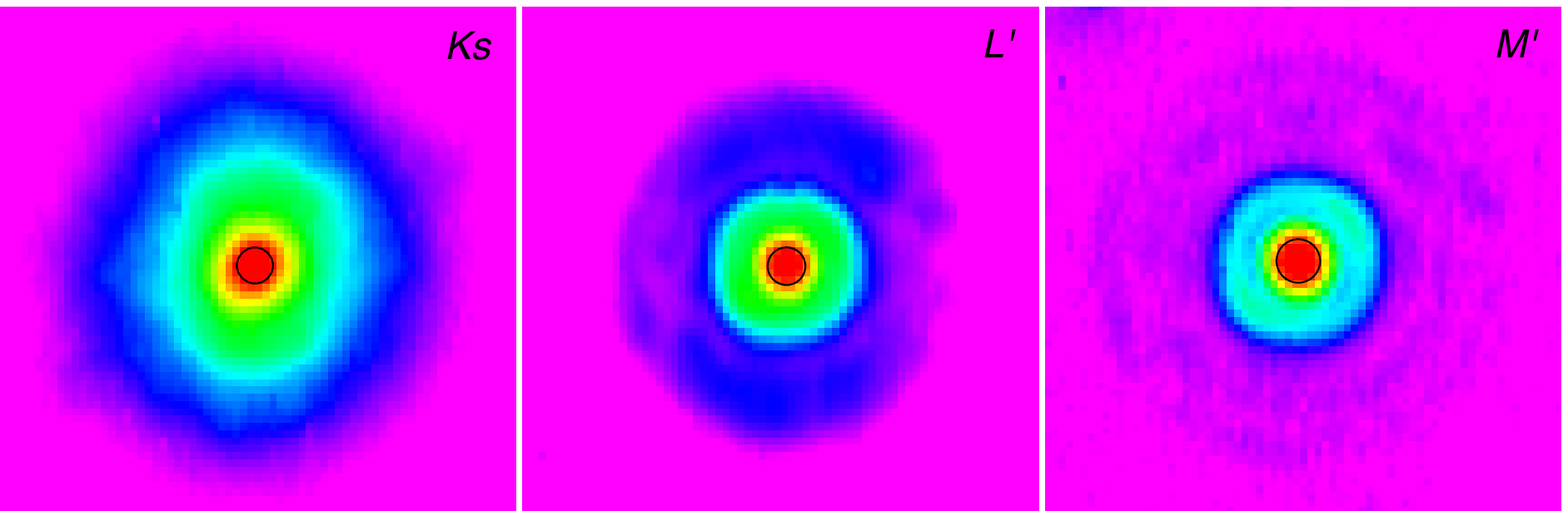}
\includegraphics[width=3.3in]{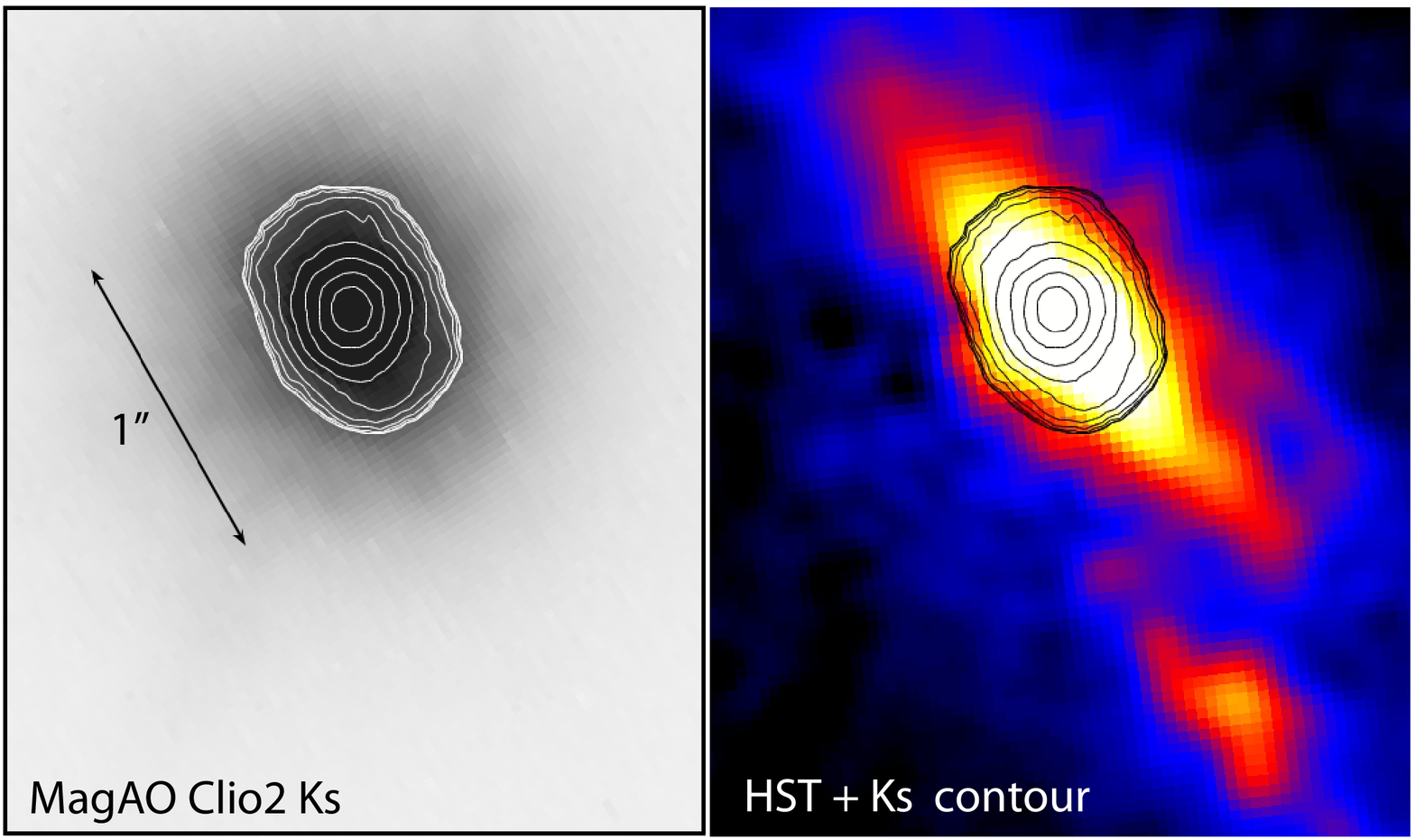}
\caption{{\it Upper panels}: MagAO $K_s$, $L^{\prime}$, and $M^{\prime}$ images of NaSt1 in false color. The images are $\approx$2\farcs7 square in angular size and oriented in detector coordinates. {\it Lower panels}: the $K_s$-band image of NaSt1 with logarithmic intensity contours (left) and the {\textit{HST}} [N~{\sc ii}] narrow-band image from Figure\,2 with the $K_s$ contours overlaid (right). An ellipsoid of extended $K_s$ emission is clearly resolved, and is approximately aligned with the major axis of the optical nebula. The lower images are oriented with north up and east toward the left.}
\label{fig:clio}
\end{figure}

Figure\,\ref{fig:clio} (lower panels) also compares the extended $K_s$ image of NaSt1 to the [N\,{\sc ii}] image from {\textit{HST}}. Interestingly, the position angle of the $K_s$ ellipsoid on the sky is roughly parallel to that of the major axis of the optical nebula, which provides further reassurance that the extended infrared structure is real. 

The FWHM of the long axis in the $K_s$ band (171\,mas) corresponds to a physical diameter of 171--513\,AU, given the range of possible distance (1--3\,kpc), and the outmost regions of the $K_s$ nebula could extend to 
700--2100\,AU. The near-infrared angular size is thus substantially larger than the 8--13\,{$\mu$m} size measured from the interferometric imaging of Rajagopal et al. (2007). Those authors found that the angular size of NaSt1 increases from 15 to 20\,mas between 8 and 13\,{$\mu$m}, respectively. This either implies that their interferometric measurements were not sensitive to lower spatial frequencies from relatively large angular sizes of $\ge$140\,mas, or that the apparent size is of NaSt1 is intrinsically larger at 2--3\,{$\mu$m} than at 8--13\,{$\mu$m}, and perhaps not the result of the same emission source.

\subsection{SINFONI spectrum}
We obtained infrared spectra of NaSt1 from the European Southern Observatory (ESO) archive. The object was observed with the ESO VLT (Yepun) telescope on 2005 Aug. 30 using the integral field spectrograph SPIFFI within the SINFONI module (Eisenhauer et al. 2003).  The instrument delivers a simultaneous three-dimensional data cube with two spatial dimensions and one spectral dimension. The $K$-band (1.95--2.45\,{$\mu$m}) grating was used with a resolving power $R\approx4000$. The pixel scale was 0\farcs1 with a field of view of 3\farcs2 square. Observations were performed under seeing-limited conditions, which were relatively poor at 1.4--1.6 arcsec. The airmass during the observations was 1.1. NaSt1 was thus spatially unresolved by SPIFFI. The data were reduced using the publicly available ESO SINFONI pipeline. The $K$-band spectrum was extracted by averaging all ``spaxels" within an aperture that encompassed the $\approx$1\farcs 5 seeing disk of the unresolved source. Telluric correction was performed using a SINFONI spectrum of the B5 star HD\,183734 from nearly the same airmass. 

 \begin{figure}
\includegraphics[width=3.4in]{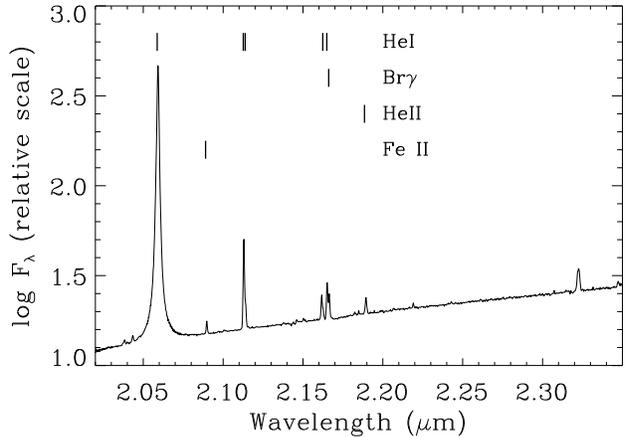}
\caption{$K$-band spectrum of NaSt1 from SINFONI.}
\label{fig:irspec}
\end{figure}

Figure\,\ref{fig:irspec} shows the $K$-band spectrum of NaSt1. The morphology and relative line strengths appear consistent with the spectrum presented in CS99. Overall, the spectrum is dominated by an extremely strong emission line of He\,{\sc i} at $\lambda$2.06\,{$\mu$m}, and the line profile exhibits Lorentzian wings extending to $\approx \pm 2500$\,km\,s$^{-1}$, similar to the H$\alpha$ line in the MMT spectrum (see Figure\,\ref{fig:spectra}). This morphology suggests that some of the emission is emerging from a region of high optical-depth, where electron scattering is occurring. The total integrated flux across the line amounts to $\approx$12\% of the integrated counts between 2.0 and 2.3\,{\micron} (before telluric correction). The spectrum also exhibits strong He\,{\sc i} emission at 2.113{$\mu$m}, a blend of Br$\gamma$ and He\,{\sc i} near $\lambda$2.166\,{$\mu$m}, He\,{\sc ii} at $\lambda$2.189{$\mu$m}, and Fe\,{\sc ii} at $\lambda$2.089\,{$\mu$m}. 

\subsubsection{The nature of the extended $K_s$-band emission}

Based on the infrared SED presented in CS99, NaSt1 is clearly a strong source of thermal infrared excess emission from dust, which is composed of at least two dust components having respective temperatures of 700\,K and 2000\,K. It is thus natural to consider this dust as a potential source of the extended flux resolved by MagAO in the $K_s$ band. However, if the extended flux is thermal continuum emission from dust, then the source should have also been resolved at $M^{\prime}$ (where NaSt1 is brightest) for any reasonable range of dust temperatures. Indeed, at $>$171\,AU distances from the central heat source, hot dust formed within an outflow should have a cooler $M^{\prime}$ counterpart with a larger angular size. Instead, the PSF centroid at $M^{\prime}$ is substantially smaller than the angular size of the resolved source in the $K_s$ band, and no such extended counterpart is detected. This raises doubts that the extended $K_s$ component results from hot dust emission. 

There is an alternative possibility that the extended $K_s$-band emission is the result of photoionization. Given the extremely bright flux of the He\,{\sc i} emission lines observed in the $K$-band spectrum, and the approximate sky alignment between the $K_s$ and [N\,{\sc ii}] extended structures, it is plausible that the MagAO observations resolved the nebula in the light of He\,{\sc i}, the flux of which is dominated by the $\lambda$2.06\,{$\mu$m} transition. This also suggests that the marginally resolved structure in the $L^{\prime}$ PSF halo could be extended He\,{\sc i} emission at $\lambda$3.703\,{$\mu$m}. The compactness of He\,{\sc i} relative to [N\,{\sc ii}] would explain why NaSt1 was not spatially resolved from the ground in the light of He\,{\sc i} by our spectroscopy (see Figure\,\ref{fig:profile}), nor by the narrowband imaging of CS99.  However, the broad Lorentzian wings in the He\,{\sc i} $\lambda$2.06\,{$\mu$m} line profile indicate that electron scattering is occurring, which is not likely to be efficient in the extended nebula at $>$171\,AU distances from the central star. Rather, this type of line profile is suggestive of emission from a dense stellar wind, which would have to be produced near the central star and, hence, in the unresolved core of the nebula. Thus, if the broad-winged He\,{\sc i} emission is responsible for the resolved flux at $K_s$, then it could be emission from the central stellar wind further scattered into the outer resolved regions by dust that is too cool to emit its own detectable continuum in the near-infrared. This could explain the larger angular size at $K_s$ versus $M^{\prime}$, since dust scattering is more efficient in the former band, as has been observed in $\eta$ Car's Homunculus nebula (Smith et al. 1998). In this case, however, it is puzzling that no optical counterpart of dust-scattered light was also detected in our {\textit{HST}} F654N optical continuum image of NaSt1.  

Unfortunately, we cannot attribute the He\,{\sc i} $\lambda$2.06\,{$\mu$m} line profile to specific regions within the nebula resolved by MagAO, since the source was spatially unresolved by the SINFONI observations. The broad-winged profile could simply be from the inner unresolved portion of the nebula, and a narrower line profile could comprise the extended nebula resolved at $K_s$. Future AO-assisted spectroscopy and near-infrared spectropolarimetry could help discriminate between direct line emission from the resolved nebula and dust scattering of stellar wind emission emerging from the unresolved interior.

We conclude that the extended flux at $K_s$ is most likely to be light from He\,{\sc i} $\lambda$2.06\,{$\mu$m}, where possibly some, and perhaps most, of the flux could have been scattered out to $>$171\,AU by dust too cool to emit strongly at $K_s$, $L^{\prime}$, and $M^{\prime}$ . Therefore, the strong point-source emission at these wavelengths must be from a compact component of hot dust near the core, perhaps physically  associated with the mid-infrared source resolved by interferometry (Rajagopal et al. 2007).

\section{Chandra X-ray observations}
We observed NaSt1 on 2014 September 24 19:01 UT using the \textit{Chandra X-ray Observatory} (Obs. ID 15700) and the Advanced CCD Imaging Spectrometer (ACIS) instrument (Weisskopf et al. 2002). A total exposure time of 12 kiloseconds resulted in the detection of a faint point source at $\alpha=\,$18:52:17.53 $\delta=\,+$00:59:44.6 (J2000.0), with an positional error of 1\farcs25 (90\% confidence), consistent with the respective optical and infrared positions of the bright central point source from {\textit{HST}} and Magellan. The proper designation for the X-ray source is thus CXOU\,J185217.53$+$0059446, in accordance with the standard naming convention for \textit{Chandra} sources.

We extracted the source photometry and spectra using the {\sc ciao} routines {\sc srcflux} and {\sc specextract}, respectively. After accounting for the background rate, the total number of counts attributable to the object is $42\pm7$ in the broad 0.5--7.0 keV energy band. The associated photon flux and energy fluxes in this band are $1.2^{0.9}_{1.6}\times10^{-5}$\,cm$^{-2}$\,s$^{-1}$ and $2.9^{2.2}_{3.7}\times10^{-14}$\,erg\,cm$^{-2}$\,s$^{-1}$, respectively (values accompanied with 90\% confidence envelopes defined by the subscript and superscript), and the average energy of the detected emission is $\langle E \rangle=1.7\pm0.3$\,keV. The fluxes within the individual soft (0.5--1.2 keV), medium (1.2--2.0 keV), and hard (2.0--7.0 keV) bands are listed in Table\,1, along with the hardness ratios, defined as $HR_{med-soft}=(F_{\rm med}-F_{\rm soft})/(F_{\rm soft}+F_{\rm med}+F_{\rm hard})$ and similarly for $HR_{hard-med}$. The hard band exhibits the largest energy flux, while the majority of counts are in the medium band between 1 and 2 keV. 

The ACIS-I energy spectrum of NaSt1 is shown in Figure\,\ref{fig:xspec}, along with an ACIS spectrum of the WN4 star WR\,18 for comparison (vertically scaled by 0.5), which we obtained from the Chandra Data Archive (Obs. ID 8910, PI: S.\,Skinner). Overall, the spectra of NaSt1 and WR\,18 appear similar, both having the majority of counts between 1 and 2\,keV, heavy absorption at energies $<1$\,keV, and comparable flux at harder energies relative to their 1--2\,keV fluxes. WR\,18 exhibits strong Si\,{\sc xiii} triplet emission at 1.84--1.86 keV; this is a common feature in the X-ray spectra of WN stars that indicates emission from a thermal plasma having a maximum line power temperature of log\,$T_{max}=7.0$\,K. NaSt1 also appears to exhibit an excess of flux at this energy; we suspect this may result from the same Si\,{\sc xv} transition, although the detection is of relatively low statistical significance. WR\,18 also exhibits a Si\,{\sc xv} feature at 2.46\,keV, which indicates the presence of an additional hot plasma component having a higher temperature (log\,$T_{max}=7.2$\,K); this line is not detected in NaSt1, although our relatively weak observational sensitivity at these higher energies does not allow us to place meaningful constraints on its presence. 

 \begin{figure}
\includegraphics[width=3.3in]{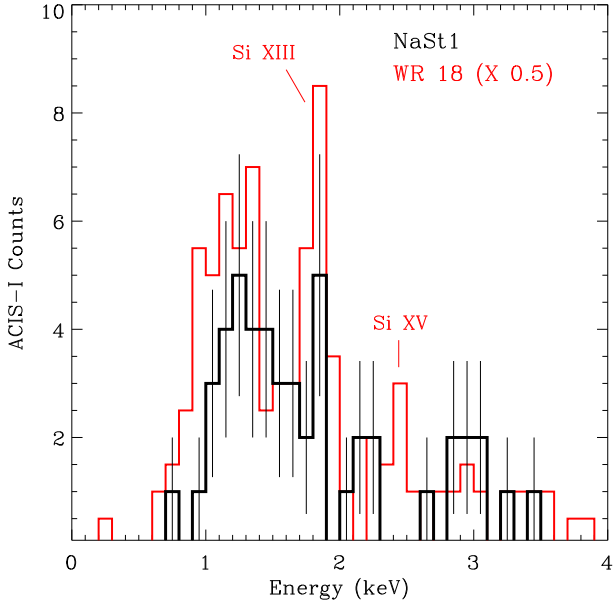}
\caption{\textit{Chandra} ACIS-I imaging spectra of NaSt1 and the WN4 star WR\,18 (red, Skinner et al. 2010) scaled by a factor of 0.5. The excess of flux for at 1.8--1.9\,keV in the spectrum of WR\,18 is due to the Si\,{\sc xiii} triplet; NaSt1 exhibits a marginally significant excess at the same energy.}
\label{fig:xspec}
\end{figure}

Skinner et al. (2010) modelled the full spectrum of WR\,18 with a two-temperature thermal plasma having $kT_{cool}=0.65\pm0.22$\,keV, $kT_{hot}=3.5\pm1.9$\,keV, and an X-ray luminosity of $L_{\rm \scriptsize X}\approx3\times10^{32}$\,erg\,cm$^{-2}$\,s$^{-1}$. Based on the observed similarities, it is plausible that the X-ray emission from NaSt1 is also emitted by a thermal plasma having physical parameters similar to WR\,18. 
Unfortunately, we are unable to attribute a unique plasma model to NaSt1 with any confidence because of the low number of counts (e.g., a power-law model can be tuned to fit the spectrum with the same level of statistical significance as a thermal plasma model, such as that used in the case of WR\,18). We can nonetheless place some constraints on the parameters of the emitting plasma using the broad-band X-ray photometry and assuming a plausible range of emission models. First, to estimate NaSt1's absorption-corrected X-ray flux, $F_{unabs}$, we adopted the extinction value of $A_V=6.5$\,mag from CS99 to estimate the total hydrogen absorption $N_H$ column along the line of sight via the $A_V$/$N_{H}$ relation from Predehl \& Schmitt (1995), and obtained $N_H=1.2\times10^{22}$\,cm$^{2}$. We applied this value to several optically thin thermal plasma models having single temperatures in the range of $kT=0.6$--2\,keV (a reasonable range for WRs), and obtained $F_{unabs}=$\,(7--38)\,$\times10^{-14}$\,erg\,cm$^{-2}$\,s$^{-1}$. We also applied a simple power-law model having a photon index in the range of $\Gamma$=1--2, and obtained a smaller range of $F_{unabs}=$\,(3--6)\,$\times10^{-14}$\,erg\,cm$^{-2}$\,s$^{-1}$. Given the likely presence of a WR in NaSt1 and the observed similarities with the X-ray properties of WR\,18, we favor a thermal plasma model for this source. If the flux we detect is mostly from a 0.65\,keV plasma component, like that of WR\,18, then the unabsorbed flux is $\approx7\times10^{-13}$\,erg\,cm$^{-2}$\,s$^{-1}$. For a distance in the range of 1--3\,kpc, the corresponding X-ray luminosity would be $L_{\rm \scriptsize X}\sim10^{32}$--$10^{33}$\,erg\,cm$^{-2}$\,s$^{-1}$. Given the potential for additional self absorption to X-rays, this should be regarded as a plausible lower limit to the X-ray luminosity. Adopting the stellar bolometric luminosity range from CS99 ($4<{\rm log}\,L/L_{\odot}<6.5$), the X-ray luminosity estimate implies $L_{\rm \tiny X}/L_{\rm \tiny bol}\sim10^{-8}$--$10^{-5}$.

 \begin{table} \setlength{\tabcolsep}{34.pt}
\renewcommand\arraystretch{1.3}
      \caption{Basic X-ray parameters for NaSt1}
  \begin{tabular}[bp]{@{}lr@{}} 
  \hline
Name & CXO\,J18521753$+$0059446   \\
$F_{\rm broad}$(0.5--7.0 keV) & $2.9_{2.2}^{3.7}\times10^{-14}$ \\ 
$F_{\rm soft}$(0.5--1.2 keV)&$4.8_{2.5}^{8.3}\times10^{-15}$ \\ 
$F_{\rm med}$(1.2--2.0 keV)&$1.1_{0.7}^{1.4}\times10^{-14}$ \\
$F_{\rm hard}$(2.0--7.0 keV)& $1.3_{0.7}^{2.1}\times10^{-14}$ \\
 $HR_{med-soft}$ & $0.15\pm0.14$\\ 
 $HR_{hard-med}$& $0.18\pm0.17$ \\  
 $\langle E \rangle$ & $1.8\pm0.3$\,keV \\ 
  $N_{\rm \tiny H}$\,(ISM) & $1.2\times10^{22}$\,cm$^{2}$\\   
 $L_{\rm \tiny X}$ & $\sim10^{32}$--$10^{33}$\,erg\,s$^{-1}$\\   
 $L_{\rm \tiny X}/L_{\rm \tiny bol}$ & $\sim10^{-8}$--$10^{-5}$\\
\hline
\hline
\hline
\end{tabular} \\
\scriptsize{{\textit{Notes.}} Energy fluxes are in units of erg\,cm$^{-2}$\,s$^{-1}$ and the superscripts and subscripts represent  the 90\% confidence envelopes. The hardness ratios ($HR_{med-soft}$) were computed using photon fluxes to calculate $(F_{\rm med}-F_{\rm soft})/(F_{\rm soft}+F_{\rm med}+F_{\rm hard})$ and similarly for $HR_{hard-med}$. The interstellar hydrogen column N$_{H}$(ISM) was computed by adopting $A_V=6.5$\,mag (CS99) and using the relation of Predehl \& Schmitt  (1995). X-ray luminosity $L_X$ was calculated using distance range of 1--3 kpc and assuming a range of intrinsic X-ray spectra properties (see text \S3). }
\end{table}

To consider the full possible extreme for self-absorption we could compare NaSt1 to the subset of WR binaries whose stellar spectra have also been completely veiled by an optically thick dust screen (e.g. the DWCL ``cocoon" stars that populate the Quintuplet Cluster in the Galactic Center; Figer et al. 1999; Mauerhan et al. 2010). Such objects are probably near the high end of self-absorption for massive stars. For example, CXOGC J174645.2$-$281547, located in the Galactic center field, is consistent with an optically thin thermal plasma having a high temperature of $kT\approx3.8$\,keV heavily absorbed by an intrinsic column of $N_{\rm \tiny H}=2\times10^{23}$\,cm$^{2}$ (Hyodo et al. 2008). If we apply such a value to NaSt1 and assume the X-rays are also produced by optically thin thermal plasma having a temperature in the broad range of 0.6--3.8 keV, then we obtain unabsorbed fluxes in the range of (1--100)\,$\times10^{-13}$\,erg\,cm$^{-2}$\,s$^{-1}$. The corresponding range of possible luminosities would therefore be $L_{\rm \scriptsize X}\sim10^{31}$--$10^{34}$\,erg\,s$^{-1}$. However, it appears that the intrinsic plasma temperature of the dominant emission component in NaSt1 is unlikely to be higher than 2 keV, given the observed drop in X-ray flux at energies above this value and the overall similarity to the cooler WR\,18. Furthermore, an absorption column as high as $N_{\rm \tiny H}=2\times10^{23}$\,cm$^{2}$ appears to be inconsistent with the data as well, since all of the emission models we have considered (optically thin thermal plasma, thermal Bremsstrahlung, and power law) would have their emission strongly suppressed at energies in the range 1--2\,keV if the absorption were this strong. Most of the X-ray flux from NaSt1 is actually in the range of 1--2\,keV, so it seems unlikely that the absorption column is greater than $10^{23}$\,cm$^{2}$. It thus appears that a total column of $N_{\rm \tiny H}=2\times10^{22}$\,cm$^{2}$ inferred from the adopted total optical-infrared extinction might be a close approximation, and could already account for a substantial fraction of the self absorption. In consideration of all of the facts, we conclude that the X-ray luminosity of NaSt1 is most likely to be $L_{\rm \scriptsize X}\sim10^{32}$--$10^{33}$\,erg\,s$^{-1}$; but given the potential for the presence of a fainter hard component that the observations were not fully sensitive to, and the potential for some additional self absorption, the actual value is probably closer to $10^{33}$\,erg\,s$^{-1}$. In turn, this implies $L_{\rm \tiny X}/L_{\rm \tiny bol}\sim10^{-7}$--$10^{-5}$. 

The X-ray properties of NaSt1 are broadly consistent with the X-ray characteristics of both purportedly single WRs and those in binaries (Skinner et al. 2010; Gagn{\'e} et al. 2012). The hot thermal plasma in these systems is generated by shocks that result from supersonic winds, in many cases by two opposing winds in collision. WR\,18, which we compared to NaSt1 in \S4, is not a confirmed binary star, so it not entirely clear if the X-ray properties in that system or in NaSt1 are the result of colliding winds. Indeed, many other colliding-wind binaries (CWBs) have more prominent hard X-ray components ($kT>2$\,keV, Gagn{\'e} et al. 2012), and in some cases exhibit  Fe K$\alpha$ emission near 6.7\,keV energy, for example, in $\eta$ Car (Corcoran et al. 2000) and WR\,147 (Skinner et al. 2007). As noted earlier, the lack of a similarly hard component in NaSt1 could simply be due to the current observational sensitivity limits. Intrinsic dimness at hard X-ray energies, on the other hand, could indicate that the winds of the component stars are relatively slow compared to the stars in more hard-X-ray bright CWBs (Luo et al. 1990; Usov 1992). Hard X-ray emission can also vary with orbital phase, so NaSt1 could have been observed in a relatively soft state. Deeper X-ray observations in the future will allow for more meaningful constraint on the presence of hard X-rays and variability in NaSt1. 

\section{DISCUSSION}
CS99 suggested that NaSt1 is a WR cloaked in an opaque dusty nebula, similar to the Homunculus around $\eta$ Car. Indeed, their nebular analysis demonstrated that NaSt1 is the only other object in the Galaxy known to exhibit such an extreme level of nitrogen enrichment (indicative of advanced CNO processing from an evolved massive star), while their derived parameters for the ionization source imply $T>>30$\,kK and log\,($L/L_{\odot}$)=5.1--6.1, consistent with a WR star. Our discovery of X-ray emission from NaSt1, shown in \S4, provides additional evidence for a WR, where hot plasma is produced by shocks in supersonic stellar winds, perhaps from colliding winds in a binary.  However, although NaSt1 shares some interesting similarities with $\eta$ Car's Homunculus, the new data highlight some obvious differences, described as follows. 

The Homunculus was created during an extremely energetic explosion that ejected at least $\approx12\,M_{\odot}$ of material at velocities of $\approx600$\,km\,s$^{-1}$ (a $\sim10^{50}$ erg event; Smith et al. 2003). The outflow exhibits a striking bipolar morphology and is a strong source of [N\,{\sc ii}] emission, but also H$\alpha$ and strong optical--infrared continuum from scattered light and thermal radiation from dust. The extended nebula around NaSt1, on the other hand, is clearly an equatorial structure, with no obvious sign of a bipolar component. The optical emission is pure [N\,{\sc ii}], with no extended H$\alpha$ emission or extended continuum evident in our {\textit{HST}} images (Figure\,1) or ground-based spectroscopy (Figure\,4). Although we have yet to obtain spatially-resolved kinematic spectroscopy of NaSt1 along the {\it minor axis} of the nebula, where we expect the outflow's apparent radial velocities to be highest, there is currently no indication that the outflow velocities are anywhere near those of $\eta$ Car's explosive Homunculus. Unfortunately, we are unable to obtain a reliable mass estimate for NaSt1's nebula based on [N\,{\sc ii}]\ alone, so information on the kinetic energy of the outflow is lacking. Nonetheless, the physical processes that created NaSt1's nebula  appear to have been far less energetic than the giant eruption that created $\eta$ Car's Homunculus, based on the available data.  Below we speculate on the origin of NaSt1 in the context of other known massive binaries that have equatorial nebulae and exhibit X-ray emission and thermal dust emission. 

\subsection{Dust formation from colliding winds?}
NaSt1's equatorial nebula and X-ray-emitting plasma, and the compact thermal infrared component detected by CS99 and resolved by Rajagopal et al. (2007), are reminiscent of a subset of WR binaries that produce dust (Gehrz \& Hackwell 1974; Williams et al. 1990; Oskinova \& Hamann 2008), episodically in some cases (Williams et al. 1978, 1990; Veen et al. 1998). The ring-like and possible spiral structure in NaSt1 brings to mind the specific cases of WR\,98a and WR\,104 (Monnier et al. 1999, 2002; Tuthill et al. 1999, 2008), which are CWBs containing WC stars surrounded by striking Archimedean spirals of hot dust emission viewed nearly face on (the so-called `pinwheel' nebulae). The dust in these systems is thought to be produced when a carbon-rich wind from a WC star mixes with the H-rich wind of a less-evolved OB companion, creating chemistry suitable for the formation of graphite or silicates. After strong shock compression, dust condenses within a conic cooling flow that develops downstream from the wind interaction zone (Williams 2008). It is interesting in this regard that the thermal infrared emission comprising the spiral nebulae of WR\,104 and WR\,98a have physical diameters of a few $\times$\,100 to 1000\,AU, because this is comparable to the infrared size we have measured for NaSt1. It thus appears plausible that the resolved structure of infrared emission revealed by our AO imaging could also be a spiral structure viewed 12$^{\circ}$ from edge-on, whether the emission is from thermal dust continuum or scattered line emission. Meanwhile, the very recent formation of the hot dust, as suggested by Rajagopal et al. (2007), suggests the possibility of new evolutionary changes that only recently created suitable conditions for dust formation. One possibility is that the central WR recently transitioned from the WN to the WC phase after the ejection of N-rich material. Indeed, there is a growing body of evidence from observations of SNe Ibn, which indicates that WN stars can undergo discrete super-Eddington eruptions before exploding (Foley et al. 2007; Pastorello et al. 2008). If this is the case for NaSt1, the eruption could have been followed by the development of a C-rich wind from the newly formed WC star.

C-rich chemistry is not a requirement for dust to form in a CWB. The circumstellar material around $\eta$ Car, for example, which is N-rich yet C-poor, exhibits firm evidence for freshly synthesized hot dust ($T=1400$--1700\,K). Similar to WC binaries, infrared monitoring has shown that $\eta$ Car forms dust episodically, apparently maximized by strong shock compression of the stellar wind during periastron encounters (Smith 2010). In this case, the post-shock cooling apparently results in the formation of corundum dust (Al$_{2}$O$_{3}$), rather than graphite or silicates. However, corundum usually produces a spectral emission feature near 12\,{$\mu$m} wavelength, which is not seen in the relatively featureless spectrum of NaSt1 presented by Smith \& Houck (2001). Silicates are also expected to produce spectral features in the 8--14\,$\mu$m range, unless the grains are extremely large. Amorphous carbon, on the other hand, can produce a featureless mid-infrared continuum (McDonald et al. 2012). Amorphous carbon has also been favored in the cases of other dust producing WRs (Williams et al. 1987), and so it might be the most plausible dust chemistry for NaSt1. However, the limited 8--14\,{$\mu$m} wavelength coverage of the existing mid-infrared spectrum could make very broad spectral features from various dust species difficult to identify, and it might be premature to characterize the mid-infrared SED as ``featureless" based on those results alone. Broader wavelength coverage of the mid-infrared spectrum will thus be necessary to elucidate the dust chemistry in NaSt1.

Another open question is whether NaSt1 forms dust episodically, similar to other WR CWBs (e.g, see Williams et al. 1992, 1994) and $\eta$ Car (Smith 2010). Such behavior would indicate an eccentric orbit for the binary. We compared the 12\,{$\mu$m} flux of NaSt1 from the {\it Wide-field Infrared Survey Explorer} ({\textit{WISE}}) {\it Allwise} catalog (Wright et al. 2010) with the flux measured at the same wavelength by Smith \& Houck (2001) over a decade earlier; no change in mid-infrared brightness was detected. However, this does not rule out episodic dust formation, which is usually evidenced by variability in the {\it near-infrared}, since the freshly synthesized dust is relatively hot. As such, mid-infrared variations are not necessarily expected, since emission at these longer wavelengths will come from cooler dust that has flowed away from the system and piled up, to become more steadily illuminated by the luminosity of the central stars.  Interestingly, van der Hucht et al. (1997) noted near-infrared variability of NaSt1 at the level of 0.6 mag over a month-long timescale, so the possibility of episodic dust formation is worth investigating with continued near-infrared monitoring in the future. Episodic dust formation might also be correlated with variations in X-ray luminosity and hardness from colliding-wind emission (e.g., Williams et al. 2013), which is also worth exploring. Such data could reveal an orbital period for the system, which would provide much needed insights into the physical parameters of the component stars.

\subsection{Is NaSt1 the product of mass transfer?}
NaSt1's nebula of slowly moving CNO-processed material could also have become an equatorial structure as the result of binary mass transfer.  If the evolved primary star recently underwent a large radial expansion, material could have been transferred to a companion via RLOF. In the case of non-conservative mass transfer, high angular-momentum material can bypass accretion and escape the system through the outer Lagrange points (Vanbeveren et al. 1998). This might explain the relatively low velocity of the outer nebular material (compared with WR winds) and its confinement to the equatorial plane. The Roche lobe need not be filled by the star itself. In relatively wide binaries mass transfer can also occur if the donor overfills it Roche lobe with a slow dense wind (WRLOF; Abate et al. 2013), and the same is probably true for dense material from a discrete mass eruption, if the material moves slowly enough. Hydrodynamic simulations of relatively low-mass RLOF binaries containing asymptotic giant branch (AGB) stars have shown that a circumbinary envelope with spiral structures and diffuse clumpy rings can form in the equatorial plane and enshroud the system (Sytov et al. 2009). We speculate that NaSt1 could be a scaled-up analog of this process for a more massive binary. 

\begin{figure}
\includegraphics[width=3.3in]{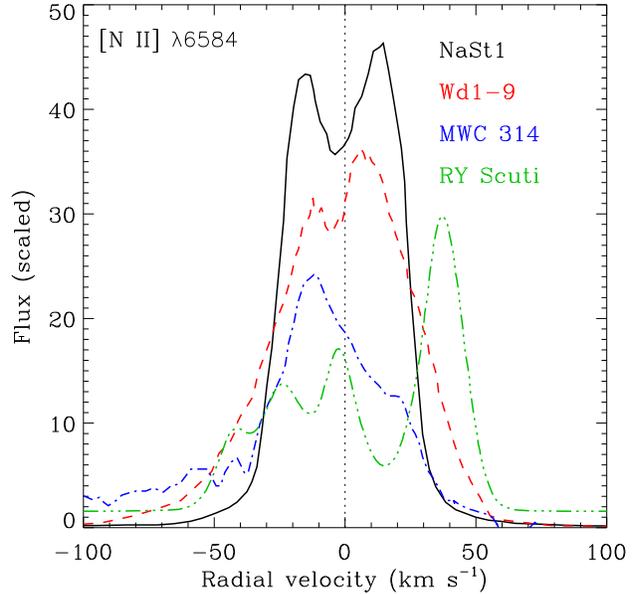}
\caption{High-resolution normalized spectrum of the [N~{\sc ii}] profile of the central point source of NaSt1, as a function of radial velocity (black solid line; reproduced from CS99), compared to that of the sgB[e]/LBV binaries Wd1-9 (red dashed line; reproduced from Clark et al. 2013 ) and MWC\,314 (blue dash-dotted line; from Lobel et al. 2013), at arbitrary orbital phases, and RY Scuti (green dash-triple-dotted line; from Smith et al. 2002). The velocities were shifted to center the entire profiles. In the cases of Wd1-9 and MWC\,314, the double peaked profiles have been interpreted as the result of rotational motion in a circumbinary disk, but could also be from an expanding ring or torus, as is the interpretation in the case of RY Scuti}
\label{fig:spec_peaks}
\end{figure}

Mass transfer might also explain some of the kinematic features previously observed in NaSt1's spectra, in particular, the double-peaked emission profiles detected by CS99. Those authors showed that the emission lines of various atomic species spanning a wide range of ionizations states exhibit an interesting correlation between profile width and ionization potential, with the highest excitation lines exhibiting the largest velocities (FWHM up to $\approx$160\,km\,s$^{-1}$). This implies that there are multiple line-forming regions in NaSt1's circumstellar envelope. These characteristics provide some important clues on the nature of NaSt1, and facilitate a comparison with other massive binaries in which mass transfer is thought to have recently occurred.

Figure\,\ref{fig:spec_peaks} shows the double-peaked [N\,{\sc ii}] emission-line profile of NaSt1 (reproduced from CS99) and for several other binary systems containing evolved massive stars (note that the spectroscopic observations which produced this profile in NaSt1 were centered on the compact central source and did not sample the larger nebula). The separation between each peak in the double profile is $\approx$30\,km\,s$^{-1}$, comparable to other systems, and is particularly similar to the object Wd1-9 from the starburst cluster Westerlund\,1 (Clark et al. 2013). Wd1-9 is composed of a supergiant B[e] star in a massive interacting binary that is undergoing, or has recently exited from, a phase of rapid RLOF. The star exhibits a rich emission-line spectrum with no photospheric features, similar to NaSt1, albeit with softer ionization\footnote{Based on the presence of He\,{\sc ii} emission, van der Hucht et al. (1997) suggested an O[e] classification for NaSt1}. In addition to the [N\,{\sc ii}] profile and overall spectral morphology, Wd1-9 is similar to NaSt1 in several other respects: it is an X-ray source having $L_X\sim10^{33}$\,erg\,s$^{-1}$ and plasma temperature consistent with colliding-wind shock emission (Clark et al. 2008); it has an equatorially concentrated mass of hot ($T\sim1000$\,K) dust having a radius of 60--600\,AU; and it exhibits a similar correlation between emission-line width and ionization potential. The presence of strong He\,{\sc ii} emission in NaSt1, however, suggests a significantly hotter ionization source than that of Wd1-9, which does not exhibit He\,{\sc ii} emission. 

Clark et al. (2013) interpreted the double-peaked line profiles of Wd1-9 as being the result of quasi-Keplerian rotation in a massive disk or torus. This interpretation has also been invoked to explain the double-peaked line profiles detected in the LBV binary MWC\,314 (Lobel et al. 2013; also shown in Figure\,\ref{fig:spec_peaks}). A rotational component from the inner regions of NaSt1's disk is thus worth considering as a potential explanation for the double-peaked profile. However, such profiles can also be produced by another mechanism.  An expanding disk, torus, or flattened hollow cone geometry can generate double-peaked line profiles, as demonstrated by Zickgraf (2003), who argued that many sgB[e] systems could actually be exhibiting a combination of rotating disk emission at relatively small radii (where permitted lines are formed) and emission from a radially outflowing disk at larger radii (where forbidden lines, like [N\,{\sc ii}], are formed). It is difficult to discriminate between these possibilities for the double-peaked profiles from the inner regions of NaSt1, although the hypothesis of quasi-Keplerian rotation might be more compatible with the observed correlation between profile width and ionization potential shown by CS99. On the other hand, the kinematics of the outer nebula, as indicated by Figure\,\ref{fig:spatial_spec}, are more likely to be the result of an equatorial outflow that is not perfectly azimuthally symmetric, since rotation at such large radii seems unlikely. 

\subsubsection{NaSt1 as an advanced analog to RY Scuti}
Another very relevant object for comparison to NaSt1 is the eclipsing binary RY Scuti. It consists of a massive O9/B0 supergiant primary currently undergoing RLOF, and an accreting secondary (likely to be an O5 star) that is buried inside an opaque accretion torus (Grundstrom et al. 2007; Smith et al. 2011b). The system as a whole is surrounded by an expanding double-ring nebula of equatorially concentrated ionized gas exhibiting multi-peaked emission profiles from both H$\alpha$ and [N\,{\sc ii}] (Smith et al. 2002), the latter of which is also shown in Figure\,\ref{fig:spec_peaks}. In this case, the multi-peaked morphology is clearly the result of non-azimuthally symmetric expansion, not rotation. The rings are $\approx$1800\,AU in extent, expanding at roughly $\pm$45\,km\,s$^{-1}$. Based on proper-motion measurements, the structure appears to have formed from multiple discrete eruptions of mass that occurred 130 and 250 yr ago. Smith et al. (2011b) suggested the possibility that the equatorial outflow was produced by an LBV-like stellar eruption that occurred in response to the sudden accretion of mass, and that the system is potentially on its way to becoming a B[e] supergiant $+$ stripped-envelope WR binary. Interpreting NaSt1 in light of RY Scuti is interesting because NaSt1 already exhibits evidence consistent with the presence of both a B[e]-like or O[e]-like supergiant star (producing the double-peaked emission-line profiles) and a hot WR (to provide the ionization and, perhaps, the X-ray emission), all surrounded by an equatorially-concentrated nebula of CNO-processed material having a physical extent over an order of magnitude larger than the ionized nebula around RY Scuti. It is thus plausible that NaSt1 is a more highly evolved analog of the RY Scuti system.

\subsection{On the lack of WR spectral features in NaSt1}
CS99 concluded that the central ionization source in NaSt1 has a stellar temperature in excess  of 30 kK, and is thus likely to be an early-type WR. Such objects typically exhibit fast winds that generate strong emission lines with FWHM velocities of $\approx2000$--3000\,km\,s$^{-1}$. The absence of any such broad spectral lines from the purported early-type WR is thus puzzling; we speculate below on several potential causes.

Heavy obscuration of the WR spectrum by dust could be exacerbated by the inclination of the system with respect to our line of sight ($\approx12^{\circ}$ from edge on). Dust formed via colliding winds or in an equatorial outflow will probably be concentrated in a similarly inclined plane, which, if sufficiently compact, could obscure the central stars, while still allowing some photoionizing radiation to escape at higher latitudes. We speculated in \S3 that the resolved structure in the $K_s$ band could be the result of He\,{\sc i} $\lambda$2.06\,{$\mu$m} emission from the central source or stellar wind scattered beyond the obscuring nebula, which would be allowable for the case of an equatorial/toroidal distribution of dust. However, if dust obscuration is to blame for the heavy suppression of the stellar spectral features, then it is difficult to explain why the same dust does not also obscure the high-ionization double-peaked emission lines detected by CS99 or the 1--2\,keV X-ray emission shown in \S4, which should arise in the hot inner regions of the system, very close to the central stars.

An alternative explanation for the lack of WR features is dilution by a more luminous companion star and the circumstellar material that produces the bright nebular spectrum. After all, WRs that form via mass-transfer can stem from stars that have significantly lower initial masses ($\ge8$\,$M_{\odot}$; Podsiadlowski, Joss, \& Tsu 1992; Vanbeveren et al. 1998) and lower luminosities than WRs that form from direct mass loss via strong winds ($\ge20$--25\,$M_{\odot}$; Meynet \& Maeder 2005; Ekstr{\"o}m et al. 2012; Smith \& Tombleson 2015).  Lower-mass WRs will likely have weaker winds (slower and less dense) than classical WRs of relatively high mass and luminosity. As such, these stars might not be easily recognizable spectroscopically as WRs, lacking similarly broad and luminous emission lines that can easily shine through dense circumstellar material, as in the case of NaSt1's nebula. Relatively slow winds were already noted as a possible explanation for the relative faintness of NaSt1 at harder X-ray energies relative to many other WR CWBs. It thus seems plausible that dilution of the WR spectrum by a luminous companion and nebula, combined with some obscuration of the WR by dust, is the reason for the lack of detectable WR features in the optical spectrum. Similarly, the infrared spectrum of the WR could be diluted by the extremely luminous ($K_s\approx6.3$\,mag) hot dust component. The $K$-band brightness of a relatively under-luminous, reddened WR at the distance of NaSt1 (adopting $M_K=-3$\,mag for faint WRs from Crowther et al. 2006) is expected to be $\approx2$--4 mag fainter than the thermal dust, even without accounting for addition circumstellar extinction, so heavy dilution of the stellar spectrum in the infrared is also plausible.

\section{Summary and Concluding remarks}
Based on the results of high-resolution optical imaging with \textit{HST}, X-ray imaging with \textit{Chandra}, and AO infrared imaging with Magellan, and the nebular analysis of CS99, we conclude that NaSt1 (WR\,122) is likely to be a WR binary enshrouded in an equatorial nebula of CNO-processed gas with a compact thermal dust component. The apparently slow kinematics and equatorial geometry of the nebula suggest that it is the byproduct of mass transfer. NaSt1 could thus be an analog to massive interacting binary systems, such as RY Scuti, MWC\,314, and Wd1-9, but at a more advanced evolutionary stage, where one of the components has lost enough mass to become a WR. NaSt1 could therefore represent a rare and important example of the recent formation of a stripped-envelope WR formed from binary interaction.

Many open questions about the NaSt1 system remain, particularly regarding the origin of various spectral features, dust chemistry, the physical parameters of the component stars, and the orbital period of the system. Such information might be obtainable via long-term photometric and monitoring and high-resolution spectroscopy in the optical and infrared. Detailed spatially-resolved chemical abundances will also shed more light on the recent evolution of the component stars and the processes that created the nebula.

\section*{Acknowledgements}
We thank the anonymous referee for an insightful review of this manuscript. The scientific results reported in this article are based on observations made with the NASA/ESA {\textit{Hubble Space Telescope}}, which is operated by the Association of Universities for Research in Astronomy, Inc., under NASA contract NAS 5-26555. These observations are associated with program HST-GO-13034. Support for program HST-GO-13034 was provided by NASA through a grant from the Space Telescope Science Institute, which is operated by the Association of Universities for Research in Astronomy, Inc., under NASA contract NAS 5-26555. The scientific results reported in this article are based on observations made by the Chandra X-ray Observatory.
Support for this work was provided by the National Aeronautics and Space Administration through Chandra Award Number GO4-15003X issued by the Chandra X-ray Observatory Center, which is operated by the Smithsonian Astrophysical Observatory for and on behalf of the National Aeronautics Space Administration under contract NAS8-03060. This research has made use of data obtained from the Chandra Data Archive and the Chandra Source Catalog, and software provided by the Chandra X-ray Center (CXC) in the application packages CIAO, ChIPS, and Sherpa. N.S. received partial support from NSF grant AST-1312221. Some observations reported here were obtained at the MMT Observatory, a joint facility of the University of Arizona and the Smithsonian Institution. This paper includes data gathered with the 6.5 meter Magellan Telescopes located at Las Campanas Observatory, Chile. This work was performed in part under contract with the California
Institute of Technology (Caltech) funded by NASA through the Sagan Fellowship Program executed by the NASA Exoplanet Science Institute. The results presented here were also based on observations made with ESO Telescopes at the La Silla Paranal Observatory under programme ID 075.B-0547(C). This publication makes use of data products from the Wide-field Infrared Survey Explorer, which is a joint project of the University of California, Los Angeles, and the Jet Propulsion Laboratory/California Institute of Technology, funded by the National Aeronautics and Space Administration.


\begin{thebibliography}{}

\bibitem[Abate et 
al.(2013)]{2013A&A...552A..26A} Abate, C., Pols, O.~R., Izzard, R.~G., Mohamed, S.~S., \& de Mink, S.~E.\ 2013, A\&A, 552, AA26 

\bibitem[Clark et 
al.(2013)]{2013A&A...560A..11C} Clark, J.~S., Ritchie, B.~W., \& Negueruela, I.\ 2013, A\&A, 560, A11 

\bibitem[Clark et 
al.(2008)]{2008A&A...477..147C} Clark, J.~S., Muno, M.~P., Negueruela, I., et al.\ 2008, A\&A, 477, 147 

\bibitem[Corcoran et al.(2000)]{2000ApJ...545..420C} Corcoran, M.~F., 
Fredericks, A.~C., Petre, R., Swank, J.~H., 
\& Drake, S.~A.\ 2000, ApJ, 545, 420 

\bibitem[Crowther 
\& Smith(1999)]{1999MNRAS.308...82C} Crowther, P.~A., \& Smith, L.~J.\ 1999, MNRAS, 308, 82 

\bibitem[Crowther et al.(2006)]{2006MNRAS.372.1407C} Crowther, P.~A., 
Hadfield, L.~J., Clark, J.~S., Negueruela, I., 
\& Vacca, W.~D.\ 2006, MNRAS, 372, 1407 

\bibitem[Davidson et al.(1986)]{1986ApJ...305..867D} Davidson, K., Dufour, 
R.~J., Walborn, N.~R., \& Gull, T.~R.\ 1986, ApJ, 305, 867 

\bibitem[Dessart et al.(2012)]{2012MNRAS.424.2139D} Dessart, L., Hillier, 
D.~J., Li, C., \& Woosley, S.\ 2012, MNRAS, 424, 2139 

\bibitem[Dufour et al.(1997)]{1997ASPC..120..255D} Dufour, R.~J., Glover, 
T.~W., Hester, J.~J., et al.\ 1997, Luminous Blue Variables: Massive Stars
in  Transition, ASPC, 120, 255, Ed. {Nota}, A. and {Lamers}, H.

\bibitem[Eisenhauer et al.(2003)]{2003SPIE.4841.1548E} Eisenhauer, F., 
Abuter, R., Bickert, K., et al.\ 2003, SPIE, 4841, 1548 

\bibitem[Ekstr{\"o}m et 
al.(2012)]{2012A&A...537A.146E} Ekstr{\"o}m, S., Georgy, C., Eggenberger, P., et al.\ 2012, A\&A, 537, A146 

\bibitem[Figer et al.(1999)]{1999ApJ...514..202F} Figer, D.~F., McLean, 
I.~S., \& Morris, M.\ 1999, ApJ, 514, 202 

\bibitem[Foley et al.(2007)]{2007ApJ...657L.105F} Foley, R.~J., Smith, N., 
Ganeshalingam, M., et al.\ 2007, ApJL, 657, L105 

\bibitem[Gagn{\'e} et al.(2012)]{2012ASPC..465..301G} Gagn{\'e}, M., Fehon, 
G., Savoy, M.~R., et al.\ 2012, Proceedings of a Scientific Meeting in 
Honor of Anthony F.~J.~Moffat, 465, 301 

\bibitem[Gehrz 
\& Hackwell(1974)]{1974ApJ...194..619G} Gehrz, R.~D., \& Hackwell, J.~A.\ 1974, ApJ, 194, 619 

\bibitem[Grundstrom et al.(2007)]{2007ApJ...667..505G} Grundstrom, E.~D., 
Gies, D.~R., Hillwig, T.~C., et al.\ 2007, ApJ, 667, 505 

\bibitem[Hachinger et al.(2012)]{2012MNRAS.422...70H} Hachinger, S., 
Mazzali, P.~A., Taubenberger, S., et al.\ 2012, MNRAS, 422, 70 

\bibitem[Hyodo et al .(2008)]{hyd08} Hyodo, Y., Tsujimoto, M., 
Koyama, K., Nishiyama, S., Nagata, T., Sakon, I., Murakami, H., 
\& Matsumoto, H.\ 2008, PASJ, 60, 173 

\bibitem[Lawrence et al.(2007)]{2007MNRAS.379.1599L} Lawrence, A., Warren, 
S.~J., Almaini, O., et al.\ 2007, MNRAS, 379, 1599 

\bibitem[Lobel et 
al.(2013)]{2013A&A...559A..16L} Lobel, A., Groh, J.~H., Martayan, C., et al.\ 2013, A\&A, 559, A16 

\bibitem[Luo et al.(1990)]{1990ApJ...362..267L} Luo, D., McCray, R., 
\& Mac Low, M.-M.\ 1990, ApJ, 362, 267 

\bibitem[Massey 
\& Conti(1983)]{1983PASP...95..440M} Massey, P., \& Conti, P.~S.\ 1983, PASP, 95, 440 

\bibitem[Mauerhan et al.(2010)]{2010ApJ...710..706M} Mauerhan, J.~C., Muno, 
M.~P., Morris, M.~R., Stolovy, S.~R., \& Cotera, A.\ 2010, ApJ, 710, 706 

\bibitem[McDonald et al.(2012)]{2012MNRAS.427..343M} McDonald, I., 
Zijlstra, A.~A., \& Boyer, M.~L.\ 2012, MNRAS, 427, 343 

\bibitem[Meynet 
\& Maeder(2005)]{2005A&A...429..581M} Meynet, G., \& Maeder, A.\ 2005, A\&A, 429, 581 

\bibitem[Monnier et al.(1999)]{1999ApJ...525L..97M} Monnier, J.~D., 
Tuthill, P.~G., \& Danchi, W.~C.\ 1999, ApJL, 525, L97 

\bibitem[Monnier et al.(2002)]{2002ApJ...567L.137M} Monnier, J.~D., 
Tuthill, P.~G., \& Danchi, W.~C.\ 2002, ApJL, 567, L137 

\bibitem[Morzinski et al.(2014)]{2014SPIE.9148E..04M} Morzinski, K.~M., 
Close, L.~M., Males, J.~R., et al.\ 2014, SPIE, 9148, 914804 

\bibitem[Nassau 
\& Stephenson(1963)]{1963LS....C04....0N} Nassau, J.~J., \& Stephenson, C.~B.\ 1963, Hamburger Sternw.~Warner \& Swasey Obs., 0 

\bibitem[Oskinova 
\& Hamann(2008)]{2008MNRAS.390L..78O} Oskinova, L.~M., \& Hamann, W.-R.\ 2008, MNRAS, 390, L78 

\bibitem[Pastorello et al.(2008)]{2008MNRAS.389..113P} Pastorello, A., 
Mattila, S., Zampieri, L., et al.\ 2008, MNRAS, 389, 113 

\bibitem[Podsiadlowski et al.(1992)]{1992ApJ...391..246P} Podsiadlowski, 
P., Joss, P.~C., \& Hsu, J.~J.~L.\ 1992, ApJ, 391, 246 

\bibitem[Predehl 
\& Schmitt(1995)]{1995A&A...293..889P} Predehl, P., \& Schmitt, J.~H.~M.~M.\ 1995, A\&A, 293, 889 

\bibitem[Rajagopal et al.(2007)]{2007ApJ...671.2017R} Rajagopal, J., Menut, 
J.-L., Wallace, D., et al.\ 2007, ApJ, 671, 2017 

\bibitem[Skinner et al.(2010)]{2010AJ....139..825S} Skinner, S.~L., Zhekov, 
S.~A., G{\"u}del, M., Schmutz, W., \& Sokal, K.~R.\ 2010, AJ, 139, 825 

\bibitem[Skinner et al.(2007)]{2007MNRAS.378.1491S} Skinner, S.~L., Zhekov, 
S.~A., G{\"u}del, M., \& Schmutz, W.\ 2007, MNRAS, 378, 1491 

\bibitem[Smith et al.(1998)]{1998AJ....116.1332S} Smith, N., Gehrz, R.~D., 
\& Krautter, J.\ 1998, AJ, 116, 1332 

\bibitem[Smith et al.(2002)]{2002ApJ...578..464S} Smith, N., Gehrz, R.~D., 
Stahl, O., Balick, B., \& Kaufer, A.\ 2002, ApJ, 578, 464 

\bibitem[Smith et al.(2003)]{2003AJ....125.1458S} Smith, N., Gehrz, R.~D., 
Hinz, P.~M., et al.\ 2003, AJ, 125, 1458 

\bibitem[Smith \& Morse(2004)]{2004ApJ...605..854S} Smith, N., \& Morse, J.~A.\ 2004, ApJ, 605, 854 

\bibitem[Smith 
\& Owocki(2006)]{2006ApJ...645L..45S} Smith, N., \& Owocki, S.~P.\ 2006, ApJL, 645, L45 

\bibitem[Smith(2010)]{2010MNRAS.402..145S} Smith, N.\ 2010, MNRAS, 402, 
145 

\bibitem[Smith et al.(2011a)]{2011MNRAS.412.1522S} Smith, N., Li, W., 
Filippenko, A.~V., \& Chornock, R.\ 2011a,  MNRAS, 412, 1522 

\bibitem[Smith et al.(2011b)]{2011MNRAS.418.1959S} Smith, N., Gehrz, R.~D., 
Campbell, R., et al.\ 2011b, MNRAS, 418, 1959 

\bibitem[Smith(2014)]{2014arXiv1402.1237S} Smith, N.\ 2014, ARA\&A, 52, 487

\bibitem[Smith 
\& Tombleson(2015)]{2015MNRAS.447..598S} Smith, N., \& Tombleson, R.\ 2015, MNRAS, 447, 598 

\bibitem[Smith 
\& Houck(2001)]{2001AJ....121.2115S} Smith, J.~D.~T., \& Houck, J.~R.\ 2001, AJ, 121, 2115 

\bibitem[Sytov et al.(2009)]{2009ARep...53..223S} Sytov, A.~Y., Bisikalo, 
D.~V., Kaigorodov, P.~V., 
\& Boyarchuk, A.~A.\ 2009, Astronomy Reports, 53, 223 

\bibitem[Tuthill et al.(1999)]{1999Natur.398..487T} Tuthill, P.~G., 
Monnier, J.~D., \& Danchi, W.~C.\ 1999, Nature, 398, 487 

\bibitem[Tuthill et al.(2008)]{2008ApJ...675..698T} Tuthill, P.~G., 
Monnier, J.~D., Lawrance, N., et al.\ 2008, ApJ, 675, 698 

\bibitem[Usov(1992)]{1992ApJ...389..635U} Usov, V.~V.\ 1992, ApJ, 389, 635 

\bibitem[Vanbeveren et 
al.(1998)]{1998A&ARv...9...63V} Vanbeveren, D., De Loore, C., \& Van Rensbergen, W.\ 1998, A\&ARv, 9, 63 

\bibitem[van der Hucht et al.(1997)]{1997ASPC..120..211V} van der Hucht, 
K.~A., Williams, P.~M., Morris, P.~W., 
\& van Genderen, A.~M.\ 1997, Luminous Blue Variables: Massive Stars in Transition, 120, 211 

\bibitem[Veen et 
al.(1998)]{1998A&A...339L..45V} Veen, P.~M., van der Hucht, K.~A., Williams, P.~M., et al.\ 1998, A\&A, 339, L45 

\bibitem[Weisskopf et al.(2002)]{2002PASP..114....1W} Weisskopf, M.~C., 
Brinkman, B., Canizares, C., et al.\ 2002, PASP, 114, 1 

\bibitem[Williams et al.(1978)]{1978MNRAS.185..467W} Williams, P.~M., 
Beattie, D.~H., Lee, T.~J., Stewart, J.~M., 
\& Antonopoulou, E.\ 1978, MNRAS, 185, 467 

\bibitem[Williams et 
al.(1987)]{1987A&A...182...91W} Williams, P.~M., van der Hucht, K.~A., \& The, P.~S.\ 1987, A\&A, 182, 91 

\bibitem[Williams et al.(1990)]{1990MNRAS.247P..18W} Williams, P.~M., van 
der Hucht, K.~A., The, P.~S., \& Bouchet, P.\ 1990, MNRAS, 247, 18P 

\bibitem[Williams et al.(1992)]{1992MNRAS.258..461W} Williams, P.~M., van 
der Hucht, K.~A., Bouchet, P., et al.\ 1992, MNRAS, 258, 461 

\bibitem[Williams et al.(1994)]{1994MNRAS.266..247W} Williams, P.~M., van 
der Hucht, K.~A., Kidger, M.~R., Geballe, T.~R., 
\& Bouchet, P.\ 1994, MNRAS, 266, 247 

\bibitem[Williams(2008)]{2008RMxAC..33...71W} Williams, P.~M.\ 2008, 
Revista Mexicana de Astronomia y Astrofisica Conference Series, 33, 71 

\bibitem[Williams et al.(2013)]{2013MNRAS.431.1160W} Williams, P.~M., Chu, 
Y.-H., Gruendl, R.~A., \& Guerrero, M.~A.\ 2013, MNRAS, 431, 1160 

\bibitem[Wright et al.(2010)]{2010AJ....140.1868W} Wright, E.~L., 
Eisenhardt, P.~R.~M., Mainzer, A.~K., et al.\ 2010, AJ, 140, 1868 

\bibitem[Zickgraf(2003)]{2003A&A...408..257Z} Zickgraf, F.-J.\ 2003, A\&A, 408, 257 

\end{thebibliography}
\end{document}